\documentclass[12pt,preprint]{aastex}

\usepackage[reqno]{amsmath}
\newcommand{\etal}{{\em et al.\ }}

\newcommand{\kms} {\hbox{km~s$^{-1}$}}

\slugcomment{Revised, resubmitted: 2008-06-04}
\shorttitle{JCMT Observations of AR6A/B}
\shortauthors{Moriarty-Schieven \etal}

\begin{document}

\title{The Local Environment of the FUor-like Objects AR~6A and 6B}

\author{Gerald H. Moriarty-Schieven}
\affil{National Research Council of Canada, Joint Astronomy Centre, 
660 N. A'ohoku Pl., Hilo, HI  96720}
\email{g.schieven@jach.hawaii.edu}

\author{Colin Aspin}
\affil{Institute for Astronomy, University of Hawaii, 640 North
A'ohoku Place, Hilo, HI 96720-2700}  
\email{caa@ifa.hawaii.edu}

\and

\author{Gary R. Davis}
\affil{Joint Astronomy Centre,  660 N. A'ohoku Pl., Hilo, HI  96720}
\email{g.davis@jach.hawaii.edu}

\clearpage

\newpage

\begin{abstract}

We present new $^{12}$CO~J=3-2 and HCN~J=3-2 molecular line maps of
the region surrounding the young star AR~6 using the 15~metre James
Clerk Maxwell Telescope.  AR~6 was previously found to be a double
source with both components exhibiting several characteristics of
FU~Orionis (FUor) eruptive variable stars.   The aims of this
investigation were to determine if the AR~6 sources are associated
with molecular outflows and if a significant reservoir of natal
molecular gas and dust exists around the stars.  These observations
form part of a large-scale study of the outflow and circumstellar
environment characteristics of FUors and FUor-like objects to place
constrains on the age and evolutionary state of sources exhibiting
FUor-like tendencies.  Our data indicates that AR~6, like FU~Orionis
itself, does not possess a CO outflow and likewise, does not show
evidence for large amounts of molecular gas in its circumstellar
environment.   In fact, AR~6 seems to lie in a local minimum of HCN
emission.  This is also found in 850~$\mu$m dust emission seen in JCMT
archival data.  We conclude that from the near-IR to the sub-mm, AR~6
is similar to FU~Orionis in several respects.  We interpret the lack
of significant dust and molecular gas in the circumstellar environment
of AR~6, together with the large near-IR thermal excess, as evidence
that the sources have exhausted their natal envelopes, that they have
at least small hot circumstellar disks, and that they are more evolved
than Class~I protostars.  This, in itself, suggests that, since FUor
eruptions have also been observed in stars with large dust mass
envelopes (e.g. V346~Nor) and with CO outflows (e.g. L1551~IRS5), FUor
events probably occur at many different stages in the early, formative
phase of a star's life, and lends support to the idea that FUor
outbursts are repetitive like their shorter-lived relatives occurring
in EXor eruptive variables. 

Finally, we study the stellar environment around AR~6 using 2MASS,
{\it Spitzer} IRAC, and {\it Chandra} ACIS images and show that, being
part of the `Spokes' young stellar cluster, AR~6 is unlike many FUors
which typically are located in more sparsely populated regions. 

\end{abstract}


\clearpage

\section{INTRODUCTION}
FUors, named after their prototype FU Orionis, are young, low-mass
protostars undergoing massive (up to 100 fold), temporary (over
periods of decades to centuries) increases in accretion producing
rates as large as $10^{-4}$ M$_{\odot}$ yr$^{-1}$ (Hartmann \& Kenyon
1985).  During these outbursts, FUors experience a brightening of up
to 6 magnitudes over a few months, followed by a much slower fading
back to their original luminosity.  For example, V1057~Cygni returned
to its pre-outburst state some 35~years after its eruption, while
FU~Orionis itself has barely declined in brightness in the 70+ years
since its discovery.  Herbig (1966, 1977) was the first researcher to
recognized FUors as young low-mass eruptive variable stars. 

Very few FUors have been observed through their outburst phase.  The
few that have are termed `classical' FUors and only about ten such
objects are currently known.  In addition to the dramatic rise in
optical brightness, typical characteristics of FUors include an
optical spectrum resembling F- or G-type supergiant stars (Herbig
1977), near-infrared (henceforth NIR) spectra similar to cooler K- or
M-type giant stars (Mould \etal 1978), and an association with bright,
often curving reflection nebulae (Goodrich 1987).  Many, but not all,
possess thermal infrared excess emission and strong sub-mm continuum
flux indicative of the presence of massive circumstellar disks and
dense, dusty envelopes (Sandell \& Weintraub 2001).  Some FUors, for
example L1551-IRS5, possess significant wide-angle molecular outflows
and collimated Herbig-Haro (HH) jets.  Such properties would suggest
that FUor events occur during the very early formative stages of a
protostar.  However, other FUors, for example FU~Orionis itself,
possess no detectable CO outflow, no obvious HH flow, and little
thermal excess and sub-mm continuum emission.  This perhaps suggests
that such FUors are older and that FUor events are repetitive in
nature occurring throughout the first few million years of a star's
life.  Some FUors have been found to be close multiple stars.
FU~Orionis has a faint, close companion $\sim$0$\farcs$5 to the south
(Wang et al. 2004; Reipurth \& Aspin 2004). Another example is is
RNO~1 where the components 1B and 1C are separated by 6$''$ and both
show FUor characteristics (Staude \& Neckel 1991; Kenyon et al. 1993).
This has led some authors to suggest that FUor eruptions are somehow
related to the presence of a close companion star, a scenario first
proposed by Bonnell \& Bastien (1992) and Clarke \& Syer (1996). 

It is clear, from the numbers of `classical' FUors found so far, that
they are quite rare events, making the discovery of new candidate
FUors clearly paramount in the study of these early stages of star
formation.  New candidate FUors, termed FUor-like objects (Reipurth \&
Aspin 1997), have been recently found, such as PP~13S (Sandell \&
Aspin 1998; Aspin \& Sandell 2001), V733~Cep (Persson's Star; Reipurth
\etal 2007), and the Braid Nebula Star (Movsessian \etal 2006).  The
term {\it FUor-like objects} is used to describe young stars that
exhibit many of the unique characteristics of `classical' FUors but
which were not observed during their eruptive (and definitive) stage.
For brevity we henceforth refer to both bona fide classical FUors and
FUor-like 
objects using the generic term FUor. Recently, Aspin \&
Reipurth (2003, henceforth AR2003) discovered a close double star
system (separation $\sim$2$\farcs$8) in which both components
exhibited several FUor characteristics.  The two stars were designated
AR~6A and 6B, and they are located near the Cone Nebula in Monoceros,
within a cluster of optical and infrared stars (Herbig 1954; AR2003).
(For convenience, we will refer in this paper to the binary as AR~6.)
The NGC 2264 complex, of which the Cone Nebula is a 
part, is an extremely active region of star formation (e.g., Lada,
Young, \& Greene 1993; Reipurth \etal 2004) and is located at a
distance of about 800~pc (Walker 1956).

The goal of this investigation is to estimate the approximate
evolutionary stage of the AR~6 FUor candidates.  A very young
protostar should still possess a large dusty envelope (Sandell \&
Weintraub 2001) and/or a molecular outflow.  We therefore used the
JCMT to map the region surrounding AR~6 in $^{12}$CO J=3-2 emission,
to look for outflow activity, and in HCN J=3-2 emission, to search for
dense gas from a circumstellar core/envelope.  In addition, we utilize
a JCMT archival SCUBA 850~$\mu$m map of the region, first published by
Wolf-Chase \etal (2003) to trace dust emission structure near AR~6.

\section{OBSERVATIONS AND DATA REDUCTION}
All data were obtained using the James Clerk Maxwell Telescope (JCMT)
on Mauna Kea, Hawaii.  $^{12}$CO J=3-2 (345~GHz) data were obtained
using RxB3, a dual polarization, single ``pixel'', single sideband
heterodyne SIS receiver operating between 315--373~GHz, during the
nights of UT 16 November, 5 December 2003, 19--20 August 2004, and 22
December 2004.  An approximately 4.5'$\times$4.5' region was observed
by raster mapping with a 7''$\times$7'' cell size, integrating for 13
seconds per point. (See Fig.~\ref{2mass-roadmap} for the
2MASS\footnote{The Two Micron All Sky Survey   (2MASS) is a joint
project of the University of Massachusetts and the   Infrared
Processing and Analysis Center/California Institute of   Technology,
funded by the National Aeronautics and Space Administration   and the
National Science Foundation.} K-band image of the region showing the
extents of the maps.)  Pointing was checked approximately every two
hours, and was found to vary by less than $\sim$2''.  The focus was
checked once or twice per night.  An RMS of 0.4~K per channel was
achieved after smoothing to 1~\kms.  The CO 3-2 channel map is
presented in Fig.~\ref{co-chan} and AR~6 is located at its origin. 

The HCN J=3-2 (266 GHz) data were obtained UT 7 March 2007 using RxA3,
a single pixel, single polarization, dual sideband SIS receiver.  A
$\sim$1.5'$\times$1.5' region was grid-mapped with 10''$\times$10''
cells. We integrated 2 minutes per point (frequency-switching),
achieving an RMS of 0.18~K per 0.26~km~s$^{-1}$ channel.
Fig.~\ref{850-hcn} shows a contour map of the integrated intensity of
HCN emission over the velocity interval 4--6~km~s$^{-1}$.   The
greyscale image shows the 850~$\mu$m dust emission in the vicinity of
AR~6, mapped using the SCUBA bolometer array (Holland \etal 1999) in
UT August and November 1999, and October 2000.  These data were first
published by Wolf-Chase \etal (2003). Data reduction and analysis are
described in that paper.

\section{RESULTS}
\subsection{The Local Environment Around AR~6}
In Fig.~\ref{2mass-roadmap} we show the K-band 2MASS image of the
region containing AR~6.  All young variable stars and AR~6 are
labeled.  As can be seen, AR~6 seems to be associated with a small
cluster of stars, some optically visible (e.g. V230~Mon, GQ~Mon, see
Fig.~2 of AR2003) and others bright only in the infrared.  It is not
possible to determine the full extent of this cluster from the 2MASS
images but it probably has many members.  We return to the matter of
the extent of this cluster in Section 3.2 below. 

Fig.~\ref{850-hcn} shows the 850~$\mu$m continuum map of Wolf-Chase
\etal (2003) overlaid with contours showing the HCN line emission in
the $\sim$1.5'$\times$1.5' survey region centered on AR~6.  In their
paper, Wolf-Chase \etal (2003) labeled several sub-mm clumps S1--S6.
Recently, Peretto, Andr{\'e} \& 
Belloche (2006) presented a map of 1.2mm continuum emission from the
same region obtained using the IRAM 30m telescope, which has a similar angular
resolution to the JCMT at those wavelengths.  Their map shows a very
similar morphology to the Wolf-Chase \etal image.  In addition,
Peretto, Andr{\'e} \& Belloche (2006) 
used a wavelet analysis to extract several new compact
sources, which are indicated on Fig.~\ref{850-hcn} as sources DMM8 -
DMM15.

We note that 850~$\mu$m and 1.2mm continuum 
trace optically thin dust emission (and hence measure the column
density of material through the cloud) while HCN~J=3-2 observations
are most sensitive to molecular gas at densities ${\sim}10^5$
cm$^{-3}$.  From Fig,~\ref{850-hcn} we see that the HCN emission
traces the S6 continuum peak reasonable well, although the S4 clump is
not detected in HCN.  In the direction of AR~6 (white cross), both the
HCN and 
850~$\mu$m/1.2mm emission are at a local minimum, and none of Peretto,
Andr{\'e} \& Belloche's (2006) compact sources lies closer to AR~6 than
$\sim$38''. At this local minimum, the
intensity of 850~$\mu$m emission is approximately 440 mJy~beam$^{-1}$
while the HCN flux is slightly less than 2~K~km~s$^{-1}$.  The HCN
flux appears to peak at three locations surrounding AR~6 with a
maximum flux of $\sim$3~K~km~s$^{-1}$.  Our conclusion therefore, is
that the circumstellar dust/gas envelope mass around AR~6 is relatively
small. 

Although the dust emission exhibits a local minimum toward AR~6, there
is some line-of-sight 850~$\mu$m emission toward AR~6 as noted above.
Assuming the emission is optically thin, we can estimate the gas mass
in this line-of-sight direction from
\begin{equation}
M = \frac{S_{\nu}d^2}{\kappa_{\nu}B_{\nu}T_d}, 
\end{equation}
where $S_{\nu}$ is the flux, $d$ is the distance, $T_d$ is the dust
temperature, and $\kappa_{\nu}$ is the mass absorption coefficient.
The value of the absorption coefficient at $\lambda = 850 {\mu}m$ was
taken to be $\kappa_{850} = 0.01$ cm$^2$ g$^{-1}$, and we assume a
gas-to-dust mass ratio of 100.  The value of $\kappa$ depends
sensitively on the properties of the dust grains (see Henning, Michel,
\& Stognienko 1995 for a review).  Taking the distance to AR~6 to be
800pc, the flux within a radius of a few arcseconds of AR~6, $S_{850}$,
measured in Janskys, is converted to mass assuming a dust temperature
$T_d$, via 
\begin{equation}
M = 4.7 \times S_{850} [exp(\frac{17K}{T_d}) - 1] \times
(\frac{\kappa_{850}}{0.01 cm^2 g^{-1}})^{-1} M_{\odot}. \end{equation}

Derived masses are very sensitive to the dust temperature.  Wolf-Chase
\etal (2003), using data from 12~$\mu$m through 850~$\mu$m, derived a
dust temperature of 23K for source S1 (Fig. \ref{850-hcn}), which
Teixeira, Zapata \& Lada (2007) identified as a dense cluster of Class
0 protostars.  Ward-Thompson \etal (2000) used 1300 through 350~$\mu$m
data to derive dust temperatures of $\sim$15K for nearby
sub-millimetre sources.  
A dust temperature of 23K yeilds $M = 2.3
M_{\odot}$, while 15K yields $M = 4.4 M_{\odot}$.  Although this is a
significant amount of material, the 
local minimum in both dust and gas implies that it is likely not
associated with the envelopes/disks around the AR~6 sources and not
available as an accretion reservoir.  Higher resolution
interferometric observations will be required to determine the extend
and dust/gas content of their envelopes and circumstellar (binary) disks.

A plot of the 850~$\mu$m dust emission, with contours of high-velocity
red-shifted (white) and blue-shifted (black) CO emission superposed,
is shown in Fig.~\ref{850-co}.  As with the HCN/850~$\mu$m map,
although 
high-velocity CO emission is found adjacent to AR~6
(Figs.~\ref{co-chan} and \ref{850-co}), in the line-of-sight toward
AR~6, the high-velocity CO emission appears to be at a local minimum.


The most prominent high-velocity features in the vicinity of AR~6 are
a red-shifted lobe centered about 45'' WSW of AR~6, and centered at
$\sim$9$^h$40$^m$56.5$^s$ and +9$^\circ$35'42''.  This lobe has a velocity
range $>$6 km s$^{-1}$ from the ambient cloud velocity
(Fig. \ref{co-chan}), and is $\sim$45'' in extent.  It appears to be
associated with an extended blue-shifted lobe, $\sim$1' in extent,
centered approximately at 9$^h$40$^m$58.6$^s$ and +9$^\circ$35'58'',
with velocity range $\sim$6 km s$^{-1}$ from the ambient cloud
velocity.  This morphology suggests a somewhat linear outflow,
oriented through the primary red lobe and the extended blue lobe, with
a position angle of about +65$^\circ$ (Fig. \ref{850-co} solid grey
line).  The origin of this outflow 
most likely lies in the 
overlap region of the main red- and blue-shifted lobes, i.e. within
about a 10-15'' radius of 6$^h$40$^m$57.7$^s$ +9$^\circ$35'52''.  
No compact 1.2mm or 850$\mu$m are found within this radius (Peretto,
Andr{\'e} \& Belloche 2006; Wolf-Chase \etal 2003; see
Fig.~\ref{850-hcn}). 
The closest 
2MASS source is
$\sim$16'' north of the 
putative site of the outflow origin.  
AR~6 is $\sim$24'' east of this position.  Neither of
these sources is likely to be the origin of this outflow.

The red-shifted clump centered at 6$^h$40$^m$58.7$^s$
+9$^\circ$36'07'' is coincident with the blue lobe discussed above,
and 
could be part of the same outflow, if that outflow is oriented close to
the 
plane of the sky.  Alternatively it may be part of another outflow, given 
the many young stars in this neighbourhood, whose blue-shifted
emission is either very weak, or is blended with the outflow discussed
above.  The axis of this outflow would be nearly north-south
(Fig. \ref{850-co} dashed
grey line).  In this case, AR~6 lies much closer to the axis of this
putative outflow, i.e. within about 10''.  However, neither red- nor
blue-shifted emission is seen toward AR~6, and so again AR~6 is
unlikly to be the origin of this emission.

Two other apparent sources of high-velocity CO emission indicate
potential outflows in this region (Figs.~\ref{co-chan} and
\ref{850-co}).  Near the south edge of the mapped region is an
apparent blue-shifted lobe, peaking at 6$^h$40$^m$52$^s$
+09$^\circ$34$'$49$''$ (J2000).  Red-shifted emission in this region
is confused (Fig.~\ref{850-co}), but in Fig.~\ref{co-chan} some
emission at velocities $>$12.5 \kms is visible, peaking at the edge of
the map, and implying an outflow position angle $\sim$-19$^\circ$
(Fig. \ref{850-co} dashed grey line).
This emission may be associated with the 850~$\mu$m ``clump''
(Fig. \ref{850-co}, unnamed in Wolf-Chase \etal (2003)) and has a peak
intensity $\sim$250mJy~beam$^{-1}$.  There are no nearby 2MASS sources
(Fig.~\ref{2mass-roadmap}). 

Much more confused is a mix of red- and blue-shifted emission
(Figs.~\ref{co-chan} and \ref{850-co}), with position angle
$\sim$90$^\circ$ (Fig. \ref{850-co} dashed grey line) and seemingly
centered on the bright 850~$\mu$m clump 
S2 (Wolf-Chase \etal 2003) or DMM13 (Peretto, Andr{\'e} \& Belloche
2006).  Several infrared stars are found in this 
vicinity, the closest being V360~Mon.  

Finally, we note that there was
little to 
no high-velocity CO emission detected in the upper half of the map. 

\subsection{The Stellar Cluster}

Fig.~\ref{mips-roadmap} shows the {\it Spitzer} MIPS 24~$\mu$m image
of the region containing AR~6.  The extent of our CO and HCN maps are
indicated by the black and white boxes, respectively.  AR~6A and 6B
are unresolved in the 24~$\mu$m image.  However, at all IRAC
wavelengths, AR~6A and  6B are (barely) resolved but AR~6A dominates
the flux from the pair.
One thing that is clear from this image is that AR~6 is certainly a
member of a young stellar cluster.  Within 2' of AR~6 are about 20
bright 24~$\mu$m sources.  This cluster was named the ``Spokes''
cluster by Teixeira et al. (2006) since many of the stars lie on
radial arms extending away from the extremely bright thermal IR source
IRAS~06382+0939 (also called NGC2264--IRAS~12 by Margulis et
al. 1989).  Teixeira et al. (2006) found IRAS~12 to be coincident with
the young binary designated RNO-West and RNO-East by Castelaz \&
Grasdalen (1988).  They determined that the two stars were a young
B-star with a low-mass companion.  This binary can be seen in
Figs.~\ref{2mass-roadmap} and \ref{mips-roadmap} near the north-east
corner of the (thin white) HCN mapping region box.  It is in the
latter figure that the Spoke-like structure of the cluster can be
readily seen.  Teixeira et al. interpreted this arrangement of young
stars as evidence that star formation was occurring in dust filaments
on a scale comparable to the Jeans length consistent with thermal
fragmentation models for cloud collapse.  We note that although AR~6
is located west-south-west of IRAS~12 and the spokes for young stars
are to the east-south-east and south-south-east, its 24$\mu$m
brightness and distance from IRAS~12 is consistent with it being a
member of the Spokes cluster.

In order to better judge the evolutionary state of AR~6 and to study
the stellar environment immediately surrounding it, we have further
analyzed the {\it Spitzer} IRAC and MIPS data of the region (NGC~2264,
PID~58, PI:Rieke) and extracted photometry in all four IRAC bands and
the MIPS 24~$\mu$m band.  This photometry, together with the
corresponding values in the NIR (taken from AR2003) are shown in
Table~\ref{photom} together with the NIR and mid-IR colors.  We have
selected number of cluster members showing either optical H$\alpha$
emission and/or variability, plus IRAS~12 for our environment study
since these are known YSOs.  Placing AR~6 in the $[3.6]-[4.5]$
vs. $[5.8]-[8.0]$, the $[3.6]-[5.8]$ vs. $[4.5]-[8.0]$, and the
$[4.5]-[5.8]$ vs. $[5.8]-[24.0]$ color-color diagrams of Gutermuth et
al. (2008, their Figs.8 and 9), we find that in all diagrams, AR~6
lies in within the area predominantly occupied by Class~II/classical
T~Tauri stars (CTTS). We note also that FU~Ori has similar IRAC colors
to AR~6.  Fig.~\ref{ccdiag} shows s $[3.6]-[4.5]$ vs. $[5.8]-[8.0]$
color-color diagram as a representative example of the ones mentioned
above.  All objects but one are Class~II sources.  The one exception
is V607~Mon which lies at the bottom-right of the Class~0/I area.
IRAS~12, the central hub of the Spokes cluster, has IRAC colors
indicative of either a Class~0/I or Class~II star. A Class~I
classification was adopted by Margulis et al. (1989).

To quantify the classification we have adopted the spectral index
defined by Lada et al. (2006) where the slope of a linear
least-squares fit defines the object class.  Specifically,
$\alpha_{IRAC}$ is the slope of the best-fit linear function to the
spectral energy distribution between 3.6 and 8.0~$\mu$m in
log$_{10}(\lambda~F_{\lambda}$) vs. log$_{10}(\lambda$) space.   The
derived values of $\alpha_{IRAC}$ for each source are shown in
Table~\ref{photom}.  For AR~6, and all but one star,
$\alpha_{IRAC}~<$0.0 indicating, by the criteria defined by Lada et
al. (2006), that the sources are of evolutionary Class~II.  The one
source where $\alpha_{IRAC}>$0.0 is V607~Mon which has a value of
+0.34.  This confirms that it is at an earlier stage of development
and is a Class~I source.   AR~6 and FU~Ori have very similar values of
$\alpha_{IRAC}$ (-0.95 and -1.15, respectively).

Finally, we have examined the {\it Chandra} ACIS-I image of the region
(NGC~2264, OBS\_ID 2540, PI~Sciortino, 100~kS) obtained from the {\it  
Chandra}
data archive.  All but one of the stars listed in Table~\ref{photom}
have {\it Chandra} detections including AR~6.  The one exception is
V350~Mon.  X-ray emission is a well-known indicator of youth in star
forming region and is generally assumed associated with magnetic
reconnection events in the star--disk interface region (Feigelson \&
Montmerle 1999).  Emission at X-ray energies have been detected in
Class~I (Broos et al. 2007 -- M~17), Class~II (Gagn\'e, Skinner, \&
Daniel 2004 -- $\rho$~Oph) and Class~III (Getman et al. 2002 --
NGC~1333) sources yet there is only a tentative detection of such  
emission from
Class~0 protostars (for non-detections see for example Giardino et  
al. 2007 -- Serpens.  For the tentative detection see Teixeira et al.  
2007 -- SMA-1 in NGC~2264).
Although X-ray emission can probe completely through molecular clouds
and detect background sources such as AGNs, the preponderance of
sources near IRAS~12 and AR~6 supports the existence of a young
stellar cluster.  We finally note that, with the spatial resolution of
the ACIS-I images ($\sim$1$''$), AR~6A and 6B (separation 2.8$''$)
should be resolved.  This is confirmed by the {\it Chandra} detection
of both stellar components of the IRAS~12 binary (separation 2.8$''$,
Castalez \& Grasdalen 1988).  However, only AR~6A is detected (152  
counts).  It is
not unexpected that AR~6A was detected since it is clearly a Class~II
CTTS.  Two possible reasons for the non-detection of AR~6B are i)  
AR~6B is younger than AR~6A and X-ray emission has not yet
switched-on.   Such a scenario would argue against the two stars are  
a physical pair since they would then not be coeval, and ii) since  
AR~6B is fainter than AR~6A, perhaps it is less massive than AR~6A  
and hence its X-ray emission is below the detection threshold of the  
ACIS data.  We note that the 1$\sigma$ "sky" noise on the 100kS ACIS- 
I image is 0.3 counts.  Taking a conservative detection limit of 5  
counts, the X-ray emission from AR~6B is at least a factor 30$\times$  
less than that from AR~6A.

\section{DISCUSSION}
\begin{figure}
 \epsscale{1.25}
 \plotone{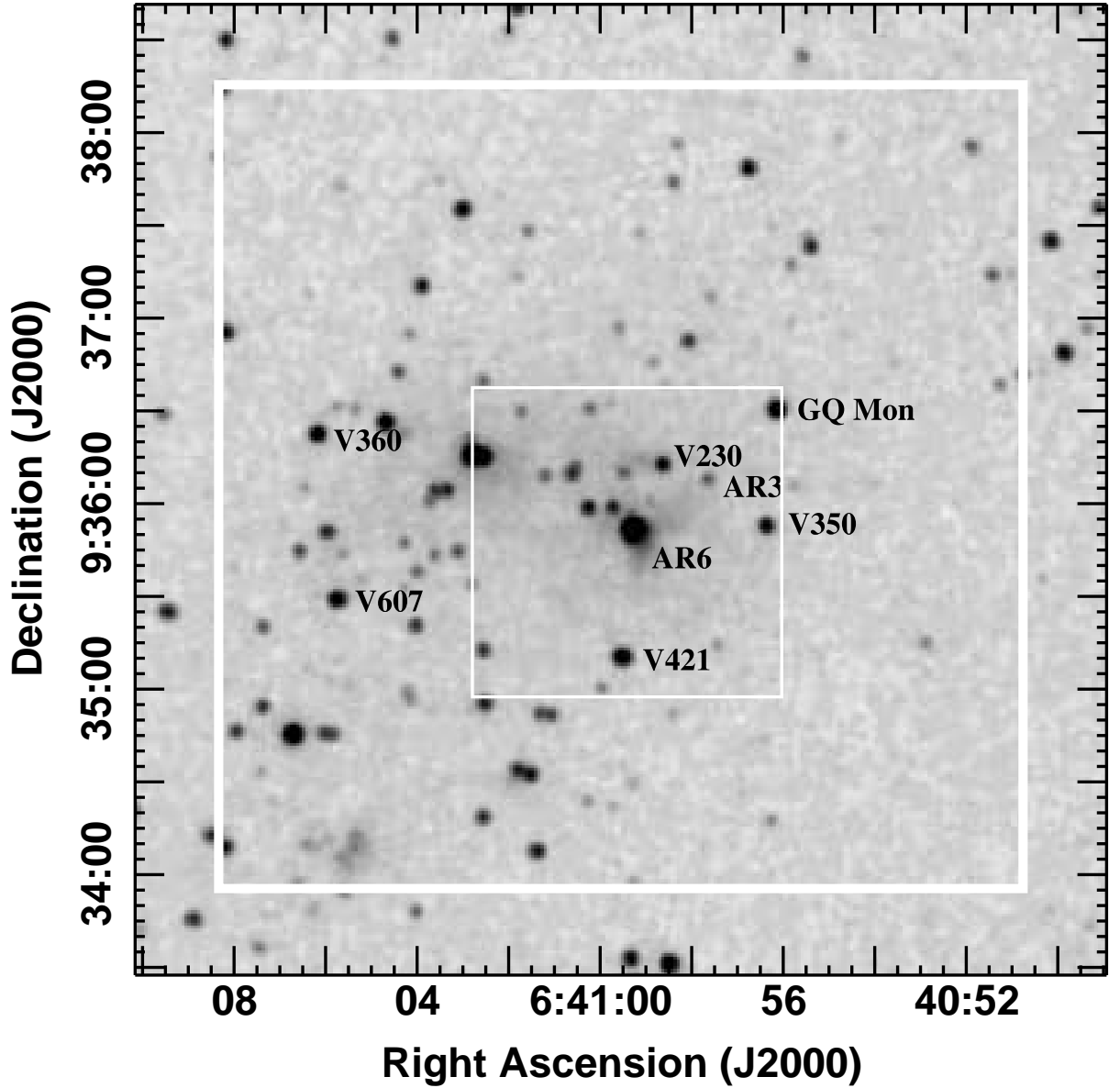}
 \caption{2MASS K-band image of the AR~6 region, showing the extent of
the regions mapped in $^{12}$CO J=3-2 (thick white box) and HCN J=3-2
(thin white box).  This entire field is encompassed by the 850 $\mu$m
emission map presented by Wolf-Chase \etal (2003). 
\label{2mass-roadmap}}
\end{figure}

\begin{figure}
 \epsscale{1.0}
 \plotone{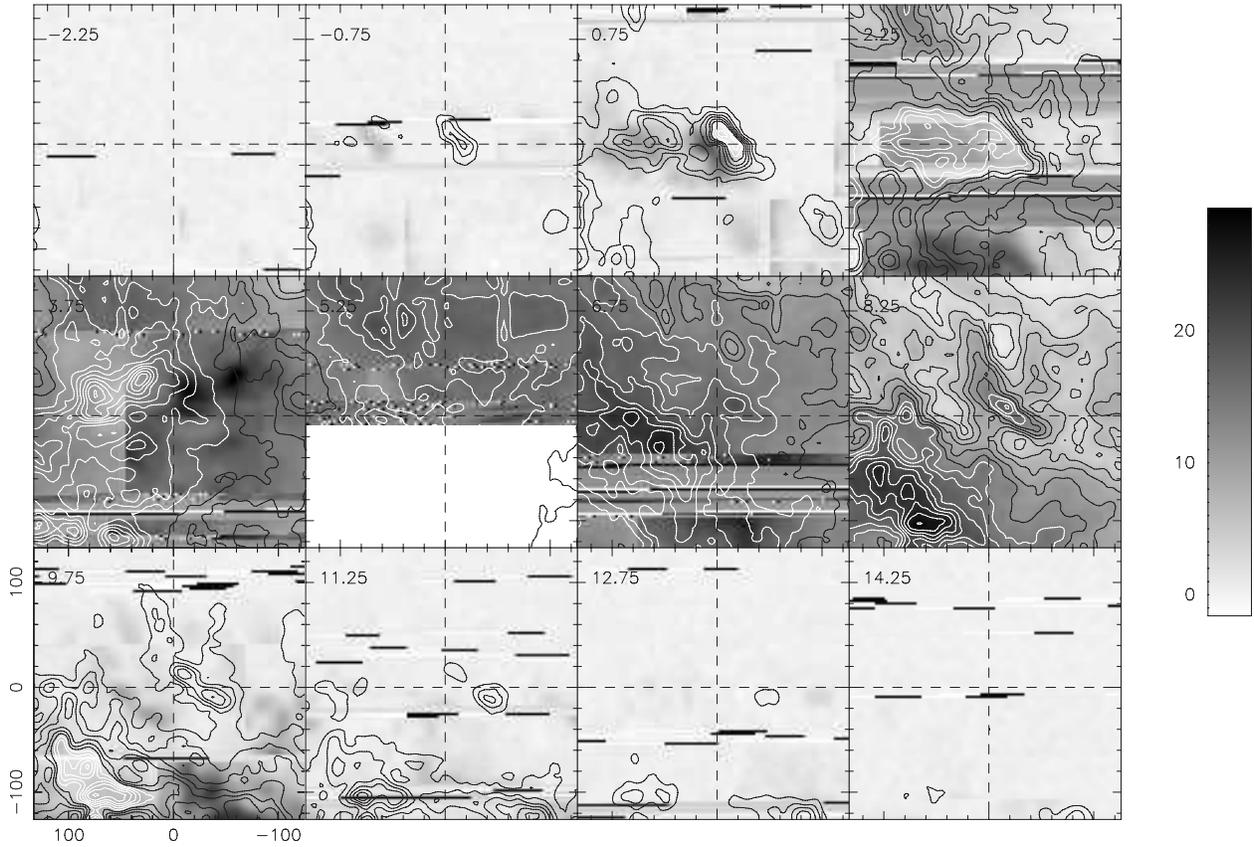}
 \caption{$^{12}$CO J=3-2 channel maps toward the AR~6 region.
Offsets are in arcseconds from the position of AR~6a (located at the
crossroads of the dashed lines in each map) at 6$^h$40$^m$59.31$^s$ 
+9$^{\circ}$35'52''.  Contours are at 2, 4, 6, ... K \kms.  Numbers in
the upper left corner of each channel map refer to the central 
velocity of the emission in the map.
\label{co-chan}} 
\end{figure}

\begin{figure}
 \epsscale{1.25}
 \plotone{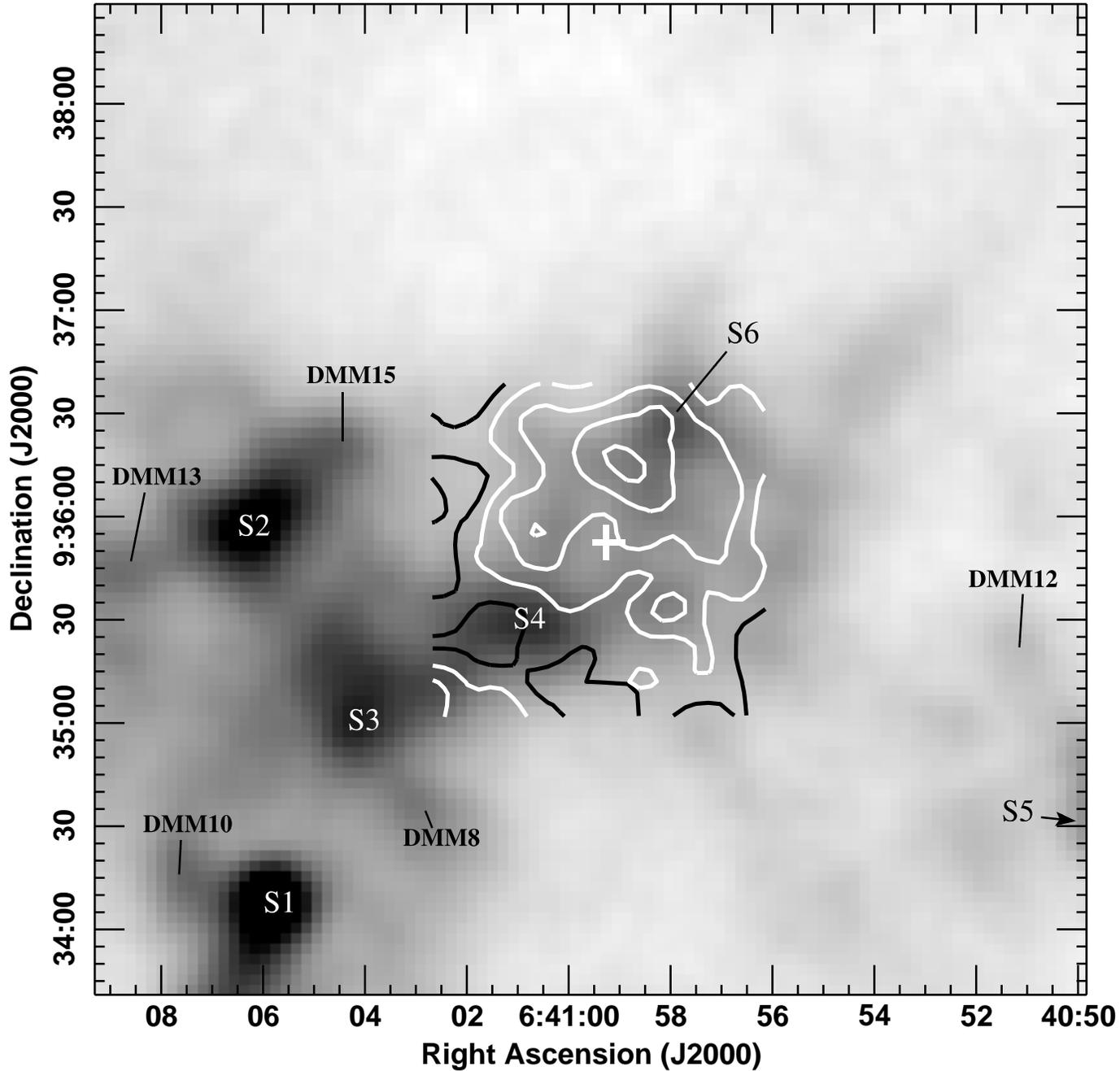}
 \caption{Greyscale image of the 850 $\mu$m
dust emission toward the AR~6 region.   The location of AR~6 is
indicated by a cross.  Contours of HCN J=3-2 integrated intensity
emission (over the velocity interval 4-7 \kms) are overlaid, with
intervals of 1 (black), 1.5, 2, 2.5, and 3 (white) K \kms.  Labeled
``clumps'' are the `12S' sources from Wolf-Chase \etal (2003) (S1 -
S6) and compact sources from Peretto, Andr{\'e} \&
Belloche (2006) (DMM8 - DMM15).
\label{850-hcn}} 
\end{figure}

\begin{figure}
 \epsscale{1.25}
 \plotone{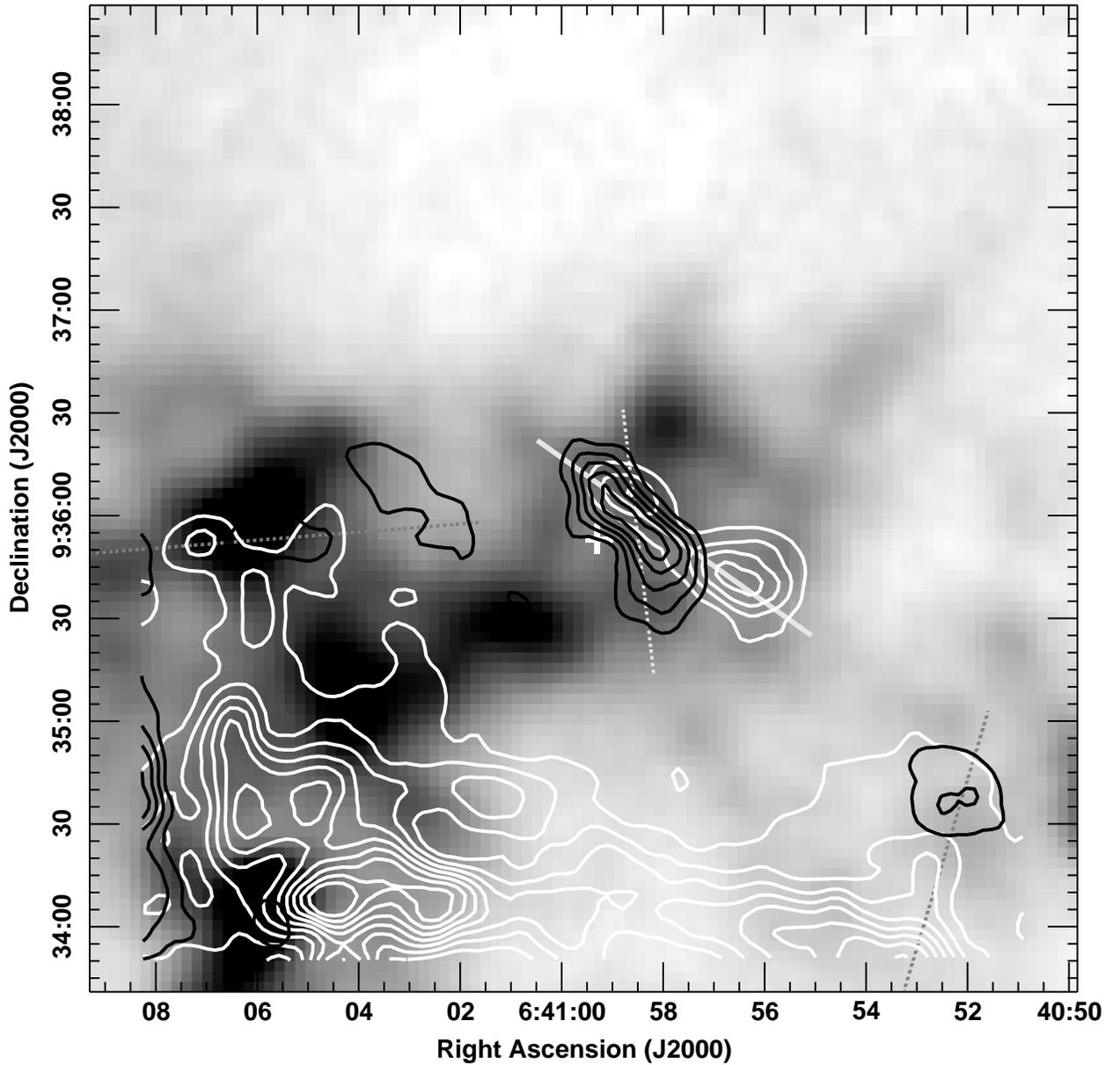}
 \caption{Same as figure \ref{850-hcn}, except the white contours
refer to red-shifted (velocity interval 9.75 to 11.75 \kms) CO, and
black contours refer to blue-shifted (velocity interval -3 to 0.5
\kms) CO. Contour intervals start at 3 K \kms and are incremented by
2~K~\kms.  The solid grey line indicates the approximate orientation
of the outflow adjacent to AR~6.  Dashed grey lines are orientations
of putative outflows discussed in the text.
\label{850-co}} 
\end{figure}

\begin{figure}
 \epsscale{1.0}
 \plotone{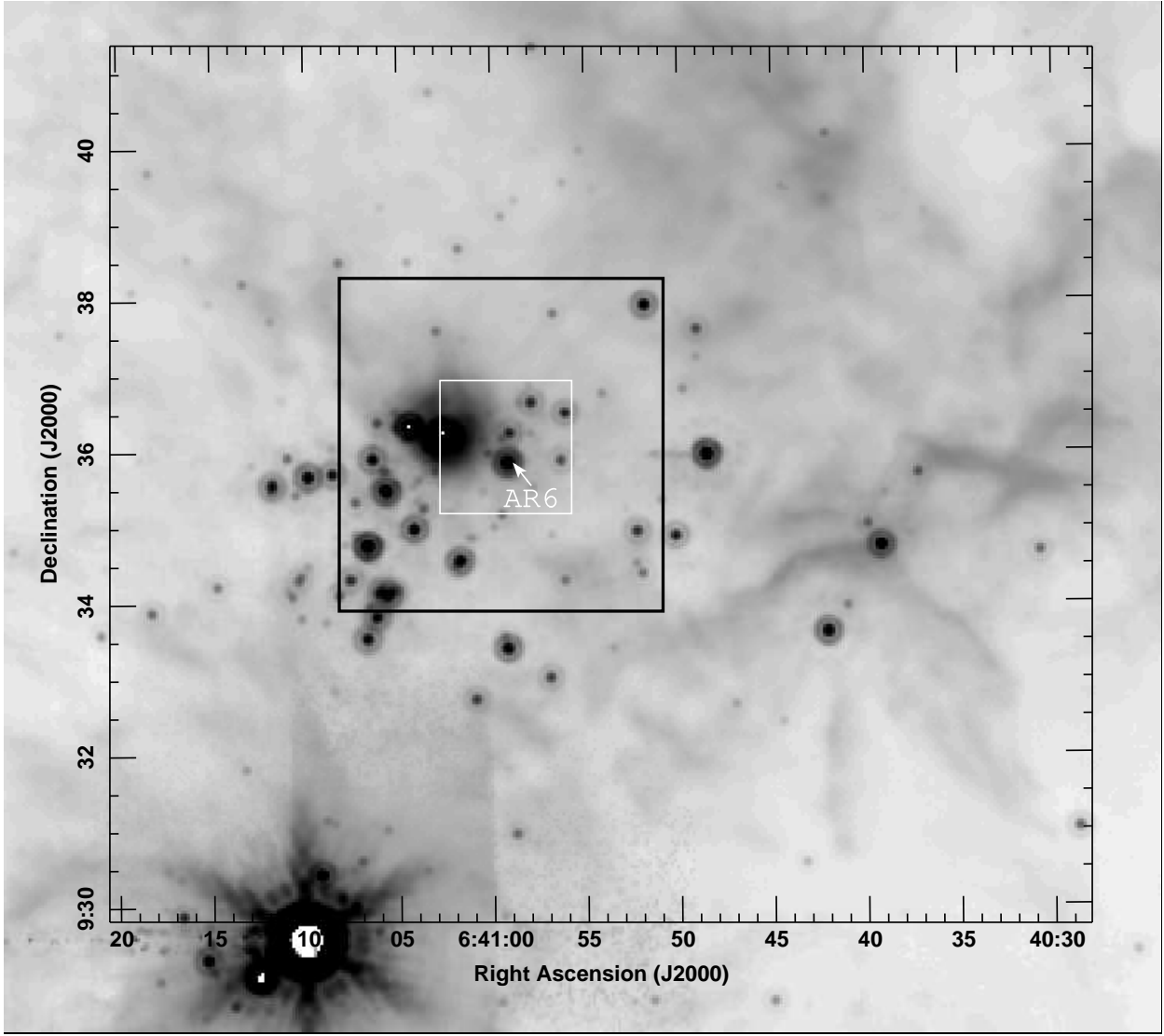}
 \caption{
Spitzer MIPS 24 $\mu$m image of the AR~6 region, 
showing the extent of
the regions mapped in $^{12}$CO J=3-2 (black box) and HCN J=3-2
(white box).  This entire field is encompassed by the 850 $\mu$m
emission map presented by Wolf-Chase \etal (2003).  Note that AR~6 is
part of the ``Spokes'' cluster discovered from Spitzer data and 
discussed by Teixeira \etal (2006, 2007).
\label{mips-roadmap}} 
\end{figure}

\begin{figure}
 \epsscale{1.0}
 \plotone{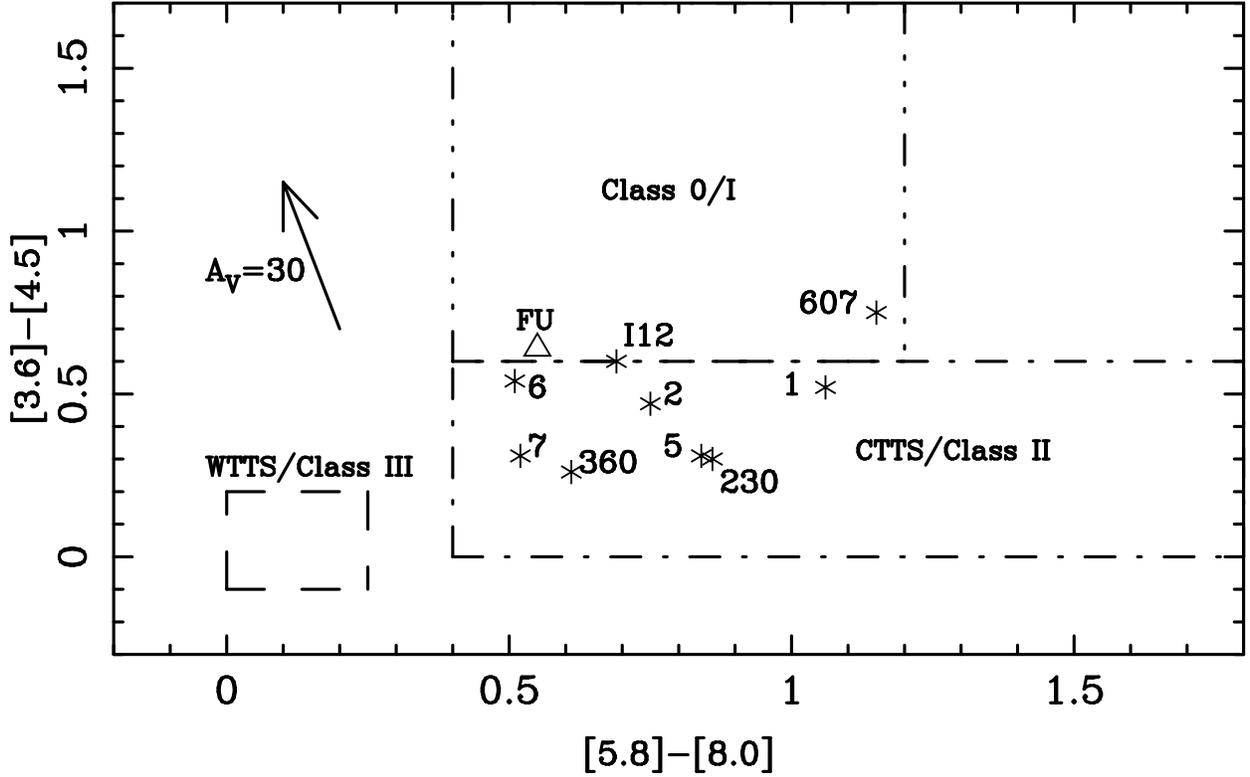}
 \caption{An IRAC color-color diagram for the sources listed in
Table~\ref{photom}. Here, we have plotted $[3.6]-[4.5]$
vs. $[5.8]-[8.0]$ and the sources are shown as stars with numeric
identifications.  These numbers correspond to either the AR number of
the source or, for those without AR identifications, the variable star
number.  'I12' refers to the IRAS~12 binary.  The open triangle
labelled 'FU' is FU~Ori.  The three rectangular regions shown are the
areas of the diagram generally populated by weak-line
T~Tauri/Class~III stars (dashed box), CTTS/Class~II stars (dot-dashed
box), and Class~0/I protostars (triple-dot-dashed box).  A vector
representing a visual extinction of A$_V$=30 magnitudes is also shown,
and was derived from the Mathis (1990) reddening law.
The x- and y-axis scales are identical to those shown in Figs.~1--4 of
Hartmann et al. (2005) in their study of {\it Spitzer} detections of
Taurus pre-main sequence stars.  In the Figures of Hartmann et al. are
displayed model colors for Class~I sources (from Allen et al. 2004)
and disk sources (from D'Alessio et al. 2005).  
\label{ccdiag}} 
\end{figure}

It is clear, from the evidence presented above and in AR2003, that the
AR~6 binary has much in common with the FUors and, in particular, with
FU~Ori itself.  In particular,

\begin{itemize}
\item AR~6 possesses curving reflection nebulosity (best seen in the
HST NIC3 image -- Fig.~3 of AR2003, but is also visible in
Fig. \ref{2mass-roadmap}) extending to the south from the
west side of AR~6. FU~Ori shows similar curving nebulosity, albeit in
the optical rather than the NIR (AR~6 -- AR2003; FU~Ori -- Goodrich
1987).   We note that several FUors show such morphological structure
in the NIR (e.g. PP~13S -- Aspin \& Sandell 2001, V733~Cep -- Reipurth
et al. 2007).

\item AR~6 is a close multiple system as is FU~Ori (AR~6 is possibly a
triple -- AR2003; FU~Ori is a binary -- Wang et al. 2004, Reipurth \&
Aspin 2004).  In addition to FU~Ori and AR~6, L1551~IRS5 (Looney,
Mundy, \& Welch 1997), RNO~1B/C (Kenyon et al. 1993), Z~CMa (Leinert
\& Haas 1987), and V1735~Cyg (Sandell \& Weintraub 2001) have all been
found to have close companions.  Bonnell \& Bastien (1992) suggested a
link between multiplicity and FUor outbursts which was also supported
by Reipurth (1989).

\item The NIR K-band spectra of both components of AR~6 and FU~Ori
show deep NIR CO overtone absorption bands and, at most, very weak
atomic absorption features.  This is a well-known characteristics of
FUors and suggests that the NIR continuum and CO overtone bandhead
absorption are formed in the disk and not in the star (AR~6 -- AR2003;
FU~Ori -- Reipurth \& Aspin 1997).  A comparison of the relative
strengths of the 2~$\mu$m CO overtone bandhead and atomic absorption
features suggests that FUors are dominated in the NIR by absorption
created in a giant- to supergiant-like atmosphere (cf. Fig.~10 from
AR2003).  FU~Ori shows an F--G~III spectrum in the optical, however,
no optical spectrum of AR~6 currently exists.

\item AR~6 has little dust emission at sub-mm/mm wavelengths and is
therefore similar to FU~Ori (AR~6 -- AR2003 and this paper; FU~Ori --
Sandell \& Weintraub 2001).  In the far-IR, however, AR~6 does not
have an IRAS association while FU~Ori is coincident with
IRAS~05426+0903 (with fluxes of 6, 14, 14, 26~Jy at 12, 25, 60, and
100~$\mu$m, respectively).   As we have seen above, AR~6 was detected
by IRAC and MIPS and has a 24~$\mu$m brightness of 1.33~Jy.  As for
other FUors, Sandell \& Weintraub (2001) found strong and extended
sub-mm/mm emission from many of the FUors observed.  Only one FUor was
not detected (BBW76).  FU~Ori itself was an weak detection
($\sim$60~mJy~beam$^{-1}$ at 850~$\mu$m) and was unresolved in their
data.  The derived dust/gas mass around FU~Ori, from isothermal model
fits, was 0.02~M$_{\odot}$, 5$\times$ smaller than the next lowest
value of 0.1~M$_{\odot}$.  The other classical FUors in the Sandell \&
Weintraub (2001) survey (e.g. V1057~Cyg, V1515~Cyg, V1735~Cyg) had
derived disk masses between 0.1 and 0.4~M$_{\odot}$.  Unfortunately,
since AR~6 does not exhibit a distinct sub-mm emission peak, we cannot
provide an estimate of circumstellar dust mass from a similar
isothermal fit.

\item AR~6, like FU~Ori, has no obvious molecular outflow (AR~6 -- this
paper; FU~Ori -- Bally \& Lada 1983; Evans et al. 1994) and appears to
lie at a local minimum in CO intensity.  FU~Ori was also observed to
be at a local minimum in the CO map of Lada \& Black (1976).  It is
interesting to note however, that Evans et al. (1994) detected
molecular outflows (using the $^{12}$CO J=3-2 line) from six of eight
FUors: specifically, RNO~1B, Z~CMa, V346~Nor, V1735~Cyg, V1515~Cyg, and
V1057~Cyg.  In addition to FU~Ori, BBW~76 was the other source in
which an outflow was not found.

\item The 2~$\mu$m spectrum of AR~6A shows relatively strong
Br$\gamma$ absorption (AR2003).  AR~6B shows no evidence of a
Br$\gamma$ feature either in absorption or emission.  Several FUors
show weak Br$\gamma$ absorption e.g. FU~Ori, V1057~Cyg, V883~Ori, and
L1551~IRS5 (Aspin, Greene, \& Reipurth 2008).  In such objects,
Br$\gamma$ absorption is likely created in either dense stellar winds
or in infalling material depending on whether it is blue- or
red-shifted (see Najita, Carr, \& Tokunaga 1996 and Muzerolle,
Hartmann, \& Calvet
1998; for further discussions and examples of Br$\gamma$
line profiles in CTTS).  In the optical, FUors often show P~Cygni
H$\alpha$ profiles with the blue-shifted absorption arising in stellar
winds.  Since we have no velocity information on the absorption, all
we can say is that both AR~6A and some FUors (including FU~Ori) show
Br$\gamma$ absorption and not emission as one might expect for a
heavily accreting young star.

\item The NIR colors of AR~6A are similar to those of FU~Ori except
that AR~6A is more heavily reddened.  The similarity of colors can be
seen in both the J-H vs. H-K and J-H vs. K-L color-color diagrams
shown in AR2003.  Both AR~6A and FU~Ori show little to no K-band
thermal excess in the J-H vs. H-K diagram yet both show a significant
thermal excess at L.  AR~6B has an enormous thermal excess at both K
and L suggesting that the circumstellar disk probably extends closer
to the stellar photosphere than in AR~6A. From the {\it Spitzer} IRAC
data, AR~6 and FU~Ori have relatively similar colors and (within the
associated uncertainties) identical spectral indices, $\alpha_{IRAC}$.
This is also true for their $[4.5]-[24.0]$ colors (4.03 and 3.98,
respectively).  All this suggests that, once reddening is removed,
AR~6 and FU~Ori have very similar intrinsic colors from 1 through to
24~$\mu$m.

\end{itemize}

We note that the new results presented here elucidate several of the
points of commonality between AR~6 and FU~Ori.  Specifically, the lack
of strong sub-mm dust and gas emission in the immediate vicinity of
the source, and the lack of active molecular outflows.

Our main conclusion from the above discussion, therefore, is that AR~6
and FU~Ori have many photometric and spectroscopic characteristics in
common and are probably in a similar evolutionary phase.  They can
both be classed as Class~II objects, and their lack of strong sub-mm
flux indicates that their circumstellar envelopes and disks are
relatively depleted.  They have likely evolved past the stage of
driving molecular outflows and HH jets (assuming they did at some
stage) and are perhaps within a few million years of the zero-age main
sequence.  Much of this is contrary to observations of other FUors
which possess large circumstellar dust masses, drive wide-angle
outflows and collimates jets, and are Class~I in nature.  The
implication, therefore, is that FUor eruptions can occur during any
phase of early stellar evolution which, in itself, hints at a
repetitive nature to these violent and definitive events.

\acknowledgements
The James Clerk Maxwell Telescope is operated by The Joint Astronomy
Centre on behalf of the Science and Technology Facilities Council of
the United Kingdom, the Netherlands Organization for Scientific
Research, and the National Research Council of Canada.  The data
presented in this paper were obtained under the following Program IDs:
m99bn07, m99bc17, m00bc32, m03bh61a, m04ad09a, and m04bc06.  This
research used the facilities of the Canadian Astronomy Data Centre
operated by the National Research Council of Canada with the support
of the Canadian Space Agency. 

This research was supported in part by NASA through the American
Astronomical Society's Small Research Grant Program.

\begin{center}
\begin{deluxetable}{lcccccccccccc}
\tabletypesize{\scriptsize}
\tablecaption{Photometry of AR~6\tablenotemark{a} and Selected Cluster
Stars\label{photom}} 
\tablewidth{0pc}
\rotate
\tablehead{
\colhead{Passband} &
\colhead{AR~6} &
\colhead{FU~Ori} &
\colhead{V350~Mon} &
\colhead{GQ~Mon} &
\colhead{V420~Mon} &
\colhead{V421~Mon} &
\colhead{V356~Mon} &
\colhead{V230~Mon} &
\colhead{V607~Mon} &
\colhead{V360~Mon} &
\colhead{IRAS~12\tablenotemark{b}} \\
\colhead{AR\#} &
\colhead{6} &
\colhead{--} &
\colhead{2} &
\colhead{1} &
\colhead{5} &
\colhead{7} &
\colhead{15} &
\colhead{--} &
\colhead{--} &
\colhead{--} &
\colhead{--} \\
\colhead{ETY\#\tablenotemark{c}} &
\colhead{--} &
\colhead{--} &
\colhead{--} &
\colhead{--} &
\colhead{--} &
\colhead{--} &
\colhead{2}  &
\colhead{--} &
\colhead{--} &
\colhead{--} &
\colhead{--} \\
\colhead{Other\tablenotemark{d}} &
\colhead{--} &
\colhead{--} &
\colhead{LH$\alpha$40} &
\colhead{LH$\alpha$38} &
\colhead{LH$\alpha$42} &
\colhead{KH$\alpha$104} &
\colhead{KH$\alpha$107} &
\colhead{--} &
\colhead{--} &
\colhead{--} &
\colhead{--} \\
\colhead{(mags)} &
\colhead{(mags)} &
\colhead{(mags)} &
\colhead{(mags)} &
\colhead{(mags)} &
\colhead{(mags)} &
\colhead{(mags)} &
\colhead{(mags)} &
\colhead{(mags)} &
\colhead{(mags)} &
\colhead{(mags)} &
\colhead{(mags)} 
}

\startdata
J                           & 11.64\tablenotemark{e} & 6.52\tablenotemark{f} & 13.20 & 12.94 & 12.71 & 11.97 & 12.91  & 12.41 & B\tablenotemark{g} & 11.64 & --   \\
H                           & 9.32                   & 5.70 & 12.12 & 11.71 & 11.96 & 11.10 & 12.83  & 11.58 & 12.73              & 11.02 & --   \\
K                           & 7.96                   & 5.16 & 11.48 & 10.77 & 11.60 & 10.72 & 12.46  & 11.13 & 10.56              & 10.55 & --   \\
L                           & 6.34                   & --   & --    & --    & --    & --    & --     & --    & --                 & --    & --   \\
M                           & 6.03                   & --   & --    & --    & --    & --    & --     & --    & --                 & --    & --   \\
$[$3.6$]$\tablenotemark{g}  & 6.42                   & 5.02 & 10.13 & 9.06  & 10.53 & 9.67  & 11.26  & 10.53 & 8.68               & 10.16 & 6.68 \\
$[$4.5$]$                   & 5.88                   & 4.38 & 9.67  & 8.53  & 10.22 & 9.37  & 10.98  & 10.23 & 7.93               & 9.90  & 6.08 \\
$[$5.8$]$                   & 5.30                   & 3.55 & 9.48  & 8.08  & 10.15 & 8.91  & 10.56  & 10.19 & 7.07               & 9.80  & 5.59 \\
$[$8.0$]$                   & 4.79                   & 3.00 & 8.73  & 7.02  & 9.31  & 8.39  & 10.89  & 9.32  & 5.92               & 9.19  & 4.90 \\
$[$24.0$]$\tablenotemark{h} & 1.85                   & 0.40 & 4.83  & 3.98  & --    & 5.64  & 6.24   & --    & 2.05               & 4.71  & 0.72 \\
J-H                         & 2.10                   & 0.82 & 0.99  & 1.23  & 0.75  & 0.87  & 0.08   & 0.83  & --                 & 0.62  & --   \\
H-K                         & 1.58                   & 0.54 & 0.73  & 0.93  & 0.36  & 0.38  & 0.37   & 0.45  & 2.17               & 0.47  & --   \\
K-L                         & 1.62                   & --   & --    & --    & --    & --    & --     & --    & --                 & --    & --   \\
$[3.6]-[4.5]$               & 0.54                   & 0.64 & 0.47  & 0.52  & 0.31  & 0.31  & 0.28   & 0.30  & 0.75               & 0.26  & 0.60 \\
$[5.8]-[8.0]$               & 0.51                   & 0.55 & 0.75  & 1.06  & 0.84  & 0.52  & --0.33 & 0.86  & 1.15               & 0.61  & 0.69 \\
$[4.5]-[5.8]$               & 0.58                   & 0.83 & 0.18  & 0.45  & 0.07  & 0.45  & 0.42   & 0.04  & 0.86               & 0.11  & 0.49 \\
$[4.5]-[24.0]$              & 4.03                   & 3.98 & 4.83  & 4.55  & --    & 3.73  & 4.75   & --    & 5.88               & 5.19  & 5.37 \\
$[3.6]-[5.8]$               & 1.12                   & 1.47 & 0.65  & 0.98  & 0.38  & 0.76  & 0.70   & 0.34  & 1.61               & 0.36  & 1.09 \\
$[4.5]-[8.0]$               & 1.09                   & 1.38 & 0.94  & 1.51  & 0.91  & 0.98  & 0.09   & 0.91  & 2.01               & 0.72  & 1.18 \\
\hline $\alpha_{IRAC}$\tablenotemark{i}              & -0.95                  & -1.15 & --1.30 & --0.53 & --1.51 & --1.34 & --2.35  & --1.53 & +0.34        & --1.77 & --0.82 \\
\enddata
\tablenotetext{a}{AR~6A is the dominant source from the optical to mid-IR.}
\tablenotetext{b}{IRAS~12 is IRAS~06382+0939 as designated by Margulis et al. (1989).}
\tablenotetext{c}{ETY\# is the identification from Young et al. (2006).}
\tablenotetext{d}{Identification from Lick and Kiso H$\alpha$ surveys.}
\tablenotetext{e}{Conservative uncertainties on all photometry and colors are $\pm$0.04 and $\pm$0.06, respectively.}
\tablenotetext{f}{JHK photometry from 2MASS.}
\tablenotetext{g}{Photometry quality flag `bad' in 2MASS catalog.}
\tablenotetext{h}{IRAC and MIPS photometry following Gutermuth et al. (2008).}
\tablenotetext{i}{IRAC spectral index using the 3.6 to 8.0$\mu$m photometry following Lada et al. (2006).}

\end{deluxetable}
\end{center}

\clearpage

\end{document}